# UV Radiation Measurement of Nitrogen in shock focusing facility

Saranyamol V. S.     Mohammed Ibrahim S.

Hypersonic Experimental Aerodynamics Laboratory (HEAL)
Department of Aerospace Engineering, Indian Institute of Technology, Kanpur,
Uttar Pradesh, India, 208016.

**ABSTRACT**

A study on UV radiation in shock focused gas has been carried out in high temperature radiating Air experimentally. Spherical shock wave focusing was achieved with the help of a contoured converging section attached to a shock tube. The measured radiation is compared with SPECAIR software to find out that the radiation corresponds to the emission of $N_2^+$ Meinal transition. The temperature of the radiating gas was estimated from SPECAIR and was found to be around 6000K.

## 1   INTRODUCTION

Shock waves are disturbances which propagates in a medium faster than the local acoustic wave speed of that medium. The large energy associated with shock waves and its ability to cause rapid change in fluid properties have enabled researchers to think and investigate the possibility of focusing a shock. The phenomenon of shock wave focusing has applications in various fields. In medical field, it is used to treat kidney stones (lithotripsy), for diamond synthesis in material science field [1], inertial confinement fusion [2], ignition techniques in engines [3], study of Richtmyer-Meshkov instability [4] etc are some of the examples. There are several methods to focus a shock wave to a point: like, using a shock tube[5], [6], exploding wire [7], micro explosive [8], electrical discharge [9] etc.

Initial research on shock wave focusing were carried out by Guderley [10], where he studied the imploding shock waves theoretically by establishing a self-similar solution for the imploding shock waves. Initial experiments for converging shock wave was carried out by Perry and Kantrowitz [11], where focusing was achieved by passing the planar shock produced in a shock tube though a tear drop shaped insert. Other converging shock experiments achieved by a shock tube are with the help of reflector inside shock tube [3], [5], annular shock tube [12], [13], contoured converging section [14] [15], [16] etc. The contoured converging section attached to the shock tube helps to reduce the diffusion losses happening to a converging shock to maximum extend. The shock dynamics theory is used to design the contoured wall for achieving this. Zhai et al. [17] designed the converging section for cylindrical shock wave focusing and Kjellander [18] designed it for spherical shock wave focusing.

Spherical shock waves are capable of focusing more energy compared to the cylindrical shocks [19], [20]. The temperature at the spherical shock focused region can reach as high as 30,000K [19], that the enclosed gas starts radiating [13], [21]. Studies on this radiating flowfield is also of great research interest since focusing a shock wave can generate dense plasma of extremely high temperatures [22]. Saito and Glass [23] did spectroscopic measurements of the radiation at the focusing point to estimate the temperature at this point. Spectroscopic measurements in a cylindrical shock focused region were made using a monochromator combined with a photomultiplier by Matsuo et al. [24] obtained by explosive shells. Spectroscopic measurements of a converging polygonal shock are carried out by Kjellander et al. [25]. All these researchers came to a conclusion that the gas at the focusing region behaves as a blackbody radiator. In our previous work, emission spectroscopy measurements of a spherical shock focused gas and the identification of the radiating species was carried out [21]. The spectrum was captured for a wavelength range of 340 to 1080nm and the emissions obtained from the measured spectra was observed. It was observed from literatures that the UV and VUV region of the spectrum contributes to 60% of the radiative heat flux in a system [26]. In view of this, in the current paper, an effort is made to analyse the emissions observed in UV region for the spherical shock focused gas. The measured spectrum is analysed with the help of radiation software SPECAIR.

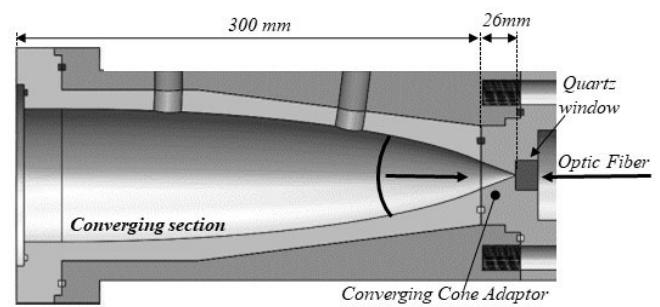

*Figure 1: Cross sectional view of the converging section.*

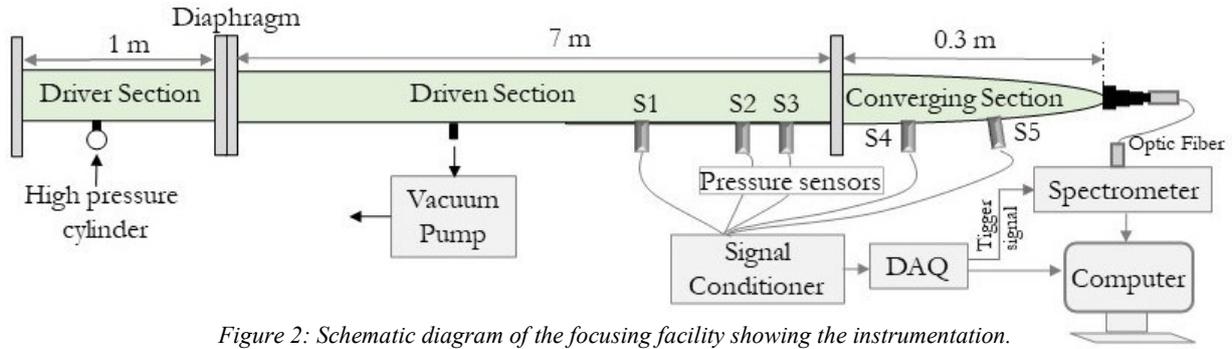

*Figure 2: Schematic diagram of the focusing facility showing the instrumentation.*

## 2 EXPERIMENTAL METHODOLOGY

The experiments are carried out in the circular shock tube facility, 'S1 (Vaigai)', at the Hypersonic Experimental Aerodynamics Laboratory (HEAL), Indian Institute of Technology, Kanpur, India. The facility has 1m long driver section, 7m long driven section and have 87 mm internal diameter. A smooth converging section is attached to the shock tube driven side end which smoothly reduces the diameter of the tube to 18 mm. The planar shock produced in the shock tube gets smoothly transformed into a spherical shock wave while passing through this converging section. The design of the converging section was made according to the geometric relations mentioned in equation 1, which was adapted from Kjellander's work [18].

$$x = A \sin\theta$$
$$y = B - R(1 - \cos\theta) \quad \quad 1$$

where $\theta$ is ranging from 0 to $0.35\pi$; A= 339.7 mm; B= 43.5 mm, and R= 47 mm.

An additional converging cone adaptor is attached to the end of the converging section which further reduces the internal diameter to 0.6 mm. The cross-sectional view of the converging section is shown in Figure 1.

Unsteady pressure sensors, ICP® pressure sensor of PCB piezotronics, model No-113B22, are located on the driven section as well as the converging section as shown in the schematic diagram (Figure 3). The pressure sensors on the shock tube are named S1, S2 and S3, and sensors on converging section are named CS1 and CS2. All sensors are connected to NI-USB6356 data acquisition system through a PCB signal conditioner (Model No. 482C05). The pressure sensors not only help to obtain the pressure variation inside the tube, but also to obtain the shock speed.

The spectrum from the radiating shocked gas is captured through an optical fibre with the help of a spectrometer attached to the shock tube as shown in Figure 3. A quartz window is placed in between the focusing point and the optical fibre to prevent damaging of the fibre. The spectrometer currently used for data acquisition is HO-CT216-3010 from Holmarc®, which captured in the UV-VIS-NIR range of 340 nm to 1080.nm. The spectrometer acquires data for 1 ms. All the data acquisition of pressure sensors as well as spectroscopic measurements are triggered by the S1 Sensor signal. The details of the setup and the acquisition methodology are explained in our previous report [21].

The measured spectrum is analysed with the help of radiation software SPECAIR. This software allows to analyse an experimentally measured spectrum. It can automatically identify the transitions in measured spectrum by numerical optimizations. The radiation of 13 radiating species of air (C, $C_2$, CN, CO, $e^-$, N, $N_2$, $N_2^+$, NH, NO, O, $O_2$, OH) and corresponding transitions can be obtained from this software.

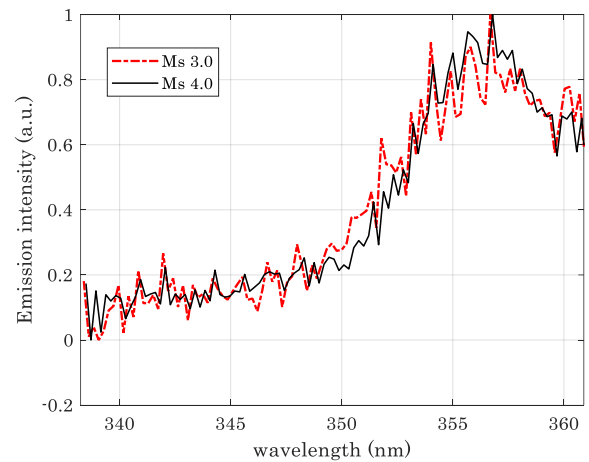

*Figure 3: Measured spectrum in UV region for Ms 3.0 and Ms 4.0.*

## 3   RESULTS AND DISCUSION

Two sets of experiments were carried out in order to study the UV emission of the radiating gas. The test gas used was Air and the driver gas used was Helium. The different initial shock Mach number achieved was $M_s$= 3.12, and 4.1. The details of test conditions are mentioned in table 1.

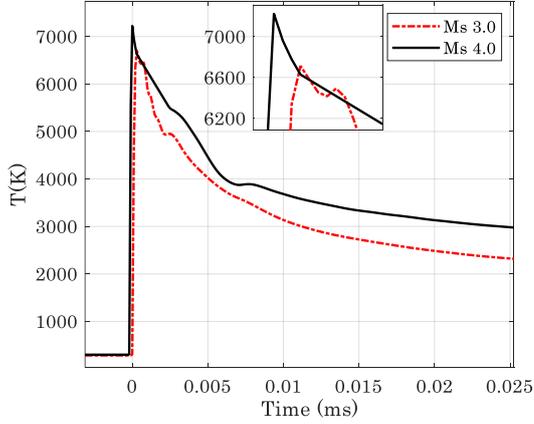

*Figure 4: The results of numerical simulations showing the temperature distribution at the focusing point*

The spectrum obtained for the two cases are shown in Figure 3. The spectra were normalised with respect to the maximum value of itself. Both cases, Ms 3.0 and 4.0, show similar spectrum, overlapping each other. To get some insights on the reason for this similar spectrum, numerical simulations were carried out with ANSYS Fluent. High-temperature effects like temperature dependent C$_P$ variation and 14 chemical reactions including three ionization reactions which includes the dissociation and recombination of species in air was included in the simulation. The detailed description of the numerical modelling is given in our previous work [27].

The simulation results shows that the temperature obtained for both cases at the focusing point has a difference of only 7% from each other. This can be seen from the numerical simulation result shown in Figure 4. Figure shows the temperature distribution monitored throughout the simulation at a point on the focusing end wall. The result suggests that even though the initial Mach number is different, the temperature obtained at the focused region is affected by the fill conditions resulting in same peak temperature at the focusing region. This is the reason for obtaining similar spectrum for both Ms 3.0 and Ms 4.0 cases.

In order to obtain the spectral analysis, the measured spectrum of MS 4.0 is given as an input to the software SPECAIR. The SPECAIR predicted emission of N$_2^+$ Meinal emissions corresponding to the measured result as shown in Figure 5. N2+ Meinal bands ( $A^2\pi_4 - X^2\Sigma_g^+$ ) emission is a common emission in shock tube experiments with air test gas [28]. The temperature prediction by SPECAIR for 4200 K and 6000K for all four temperatures: translational, rotational, vibrational, and electronic temperature is shown here. The value of 6000 K was finalised after several iterations in the software owing to the best fit to the measured spectrum. The temperature in the software was varied individually to get a perfect match with the measured spectrum. and the effect of electronic, vibration, translational and rotational temperatures were analysed. The result is shown in Figure 6.

*Table 1: Test conditions*

| $M_S$ | Driven gas Pressure | Nomenclature |
|---|---|---|
| 3.12 | 0.025 MPa | $M_S$ 3.0 |
| 4.1 | 0.01 MPa | $M_S$ 4.0 |

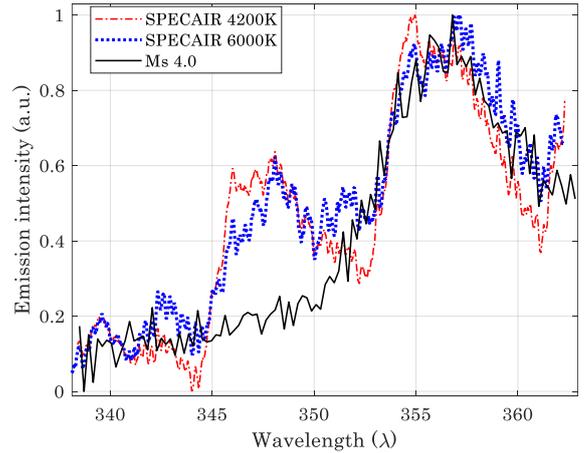

*Figure 5: SPECAIR predicted spectrum for UV emission*

Keeping all the other three temperatures as equal to 6000 K, one temperature was varied to obtain the effect of that particular temperature on the spectrum. In Figure 6 the translation temperature above 4000 K has least effect on the spectrum variation. However, with change in rotational temperature and vibration temperature, the spectrum changes its emission intensity. The electronic temperature also shows no influence on the spectrum for temperatures above 600 K. This shows that the translational and electronic temperature has least effect on the radiation.

After several iterations in the temperature loop, the optimum values for the temperatures were obtained from SPECAIR. Translational temperature is 6000 ±500 K, vibrational temperature is 6000±500K, rotational temperature is 7000 ±500K and electronic temperature is 600±50K. The final spectrum obtained is shown in Figure 7.

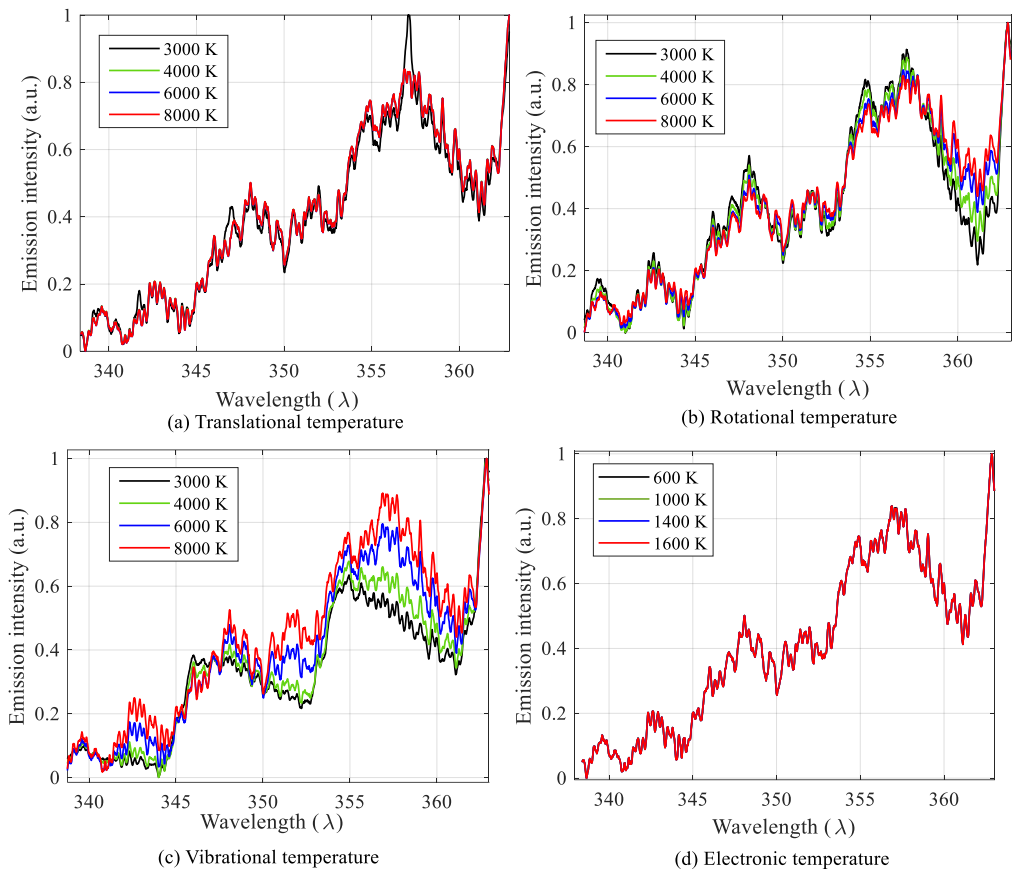

Figure 6: Effect of temperature of the prediction of spectrum by SPECAIR

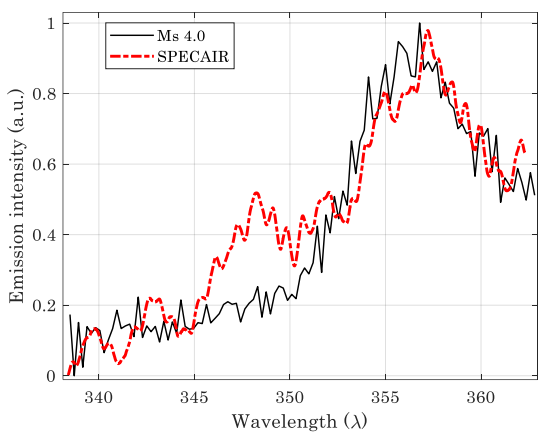

Figure 7: Comparison of SPECAIR predicted spectrum and measured spectrum

The emission at 348 nm is not observed in the measured spectrum. It is important to note here that the lifetime of an emission is important while analysing a spectrum [29]. The lifetime of $N_2^+$ $A^2\pi_4$ state is about 10 μs [30]. This could be the reason to miss one emission of $N_2^+$ at 348 nm.

## 4 CONCLUSION

Experiments were carried out to study the UV radiation in a shock focused air. The radiation observed for Ms 3.0 and Ms 4.0 was same owing to same temperature observed at the focusing region for both the cases. Numerical simulations with ANSYS fluent gave an initial insight on the temperature at the focusing region. The spectrum was further analysed with SPECAIR software to find the species transition. It was seen that the emission of $N_2^+$ Meinal band ($A^2\pi_4 - X^2\Sigma_g^+$) was emitting in this region. The temperature of the radiating gas was also obtained from the SPECAIR software and was found to be ~ 6000K.

## ACKNOWLEDGMENTS

The authors would like to thank the Science and Engineering Research Board (SERB), India, for supporting this research work under the Early Career Research Award, ECRA/2018/000678.

Reference

[1] I. I. Glass and S. P. Sharma, "Production of diamonds from graphite using explosive-driven implosions," *AIAA Journal*, vol. 14, no. 3, pp. 402–404, Mar. 1976.

[2] J. H. Nuckolls and L. L. Wood, "Inertial Confinement Fusion," *Energy in Physics, War and Peace*. pp. 157–158, 1988. doi: 10.1007/978-94-009-3031-5_10.

[3] B. Zhang, Y. Li, and H. Liu, "Analysis of the ignition induced by shock wave focusing equipped with conical and hemispherical reflectors," *Combust Flame*, vol. 236, Feb. 2022, doi: 10.1016/J.COMBUSTFLAME.2021.111763.

[4] Z. Zhai, L. Zou, Q. Wu, and X. Luo, "Review of experimental Richtmyer–Meshkov instability in shock tube: From simple to complex:," *Proc. IMech. E. Part C:J. Mechanical Engineering Science*, vol. 232, no. 16, pp. 2830–2849, Aug. 2017, doi: 10.1177/0954406217727305.

[5] B. Sturtevant and V. A. Kulkarny, "The focusing of weak shock waves," *J Fluid Mech*, vol. 73, no. 4, pp. 651–671, 1976.

[6] V. Eliasson, N. Tillmark, A. J. Szeri, and N. Apazidis, "Light emission during shock wave focusing in air and argon," *Phys. Fluids*, vol. 19, p. 106106, 2007, doi: 10.1063/1.2796214.

[7] R. S. Dennen and L. N. Wilson, "Electrical Generation of Imploding Shock Waves," *Exploding Wires*, pp. 145–157, 1962, doi: 10.1007/978-1-4684-7505-0_12.

[8] "Implosion of a spherical shock wave reflected from a spherical wall," *J. Fluid Mech*, vol. 530, pp. 223–239, 2005, doi: 10.1017/S0022112005003587.

[9] G. Gustafsson, "Experiments on shock-wave focusing in an elliptical cavity," *J Appl Phys*, vol. 61, no. 11, pp. 5193–5195, 1987, doi: 10.1063/1.338300.

[10] G. Guderley, "Strong spherical and cylindrical compression shocks near the center point of the sphere or the cylinder axis.," *Aviation Research*, vol. 19, no. 302, 1942.

[11] R. W. Perry and A. Kantrowitz, "The production and stability of converging shock waves," *J Appl Phys*, vol. 22, no. 7, pp. 878–886, 1951, doi: 10.1063/1.1700067.

[12] V. Eliasson, "The production of converging polygonal shock waves by means of reflectors and cylindrical obstacles," *AIP Conf Proc*, vol. 832, pp. 445–449, 2006, doi: 10.1063/1.2204539.

[13] V. Eliasson, "On focusing of shock waves," PhD Thesis, Royal Institute of Technology, Sweden, 2007.

[14] Z. Zhai, C. Liu, F. Qin, J. Yang, and X. Luo, "Generation of cylindrical converging shock waves based on shock dynamics theory," *Physics of Fluids*, vol. 22, no. 4, pp. 1–3, 2010, doi: 10.1063/1.3392603.

[15] X. Luo, T. Si, J. Yang, and Z. Zhai, "A cylindrical converging shock tube for shock-interface studies," *Review of Scientific Instruments*, vol. 85, no. 1, p. 015107, Jan. 2014, doi: 10.1063/1.4861357.

[16] M. Kjellander, N. Tillmark, and N. Apazidis, "Energy concentration by spherical converging shocks generated in a shock tube," *Physics of Fluids*, vol. 24, no. 126103, 2012.

[17] Z. Zhai *et al.*, "Parametric study of cylindrical converging shock waves generated based on shock dynamics theory," *Physics of Fluids*, vol. 24, no. 2, 2012, doi: 10.1063/1.3682376.

[18] M. Kjellander, "Energy concentration by converging shock waves in gases," PhD Thesis, Royal Institute of Technology, Sweden, 2012.

[19] M. Liverts and N. Apazidis, "Limiting Temperatures of Spherical Shock Wave Implosion," *Phys Rev Lett*, vol. 116, no. 014501, 2016.

[20] V. S. Saranyamol, N. Soumya Ranjan, and S. Mohammed Ibrahim, "Numerical Study of Spherical and Cylindrical Shock Wave Focusing," *Green Energy and Technology*, pp. 15–30, 2021, doi: 10.1007/978-981-15-5667-8_2.

[21] V. S. Saranyamol and S. Mohammed Ibrahim, "Effect of shock strength on the radiation of focusing shock wave," *European Journal of Mechanics, B/Fluids*, vol. 97, pp. 128–135, Jan. 2023, doi: 10.1016/J.EUROMECHFLU.2022.10.001.

[22] R. Knystautas and J. H. Lee, "Experiments on the stability of converging cylindrical detonations," *Combust Flame*, vol. 16, no. 1, pp. 61–73, 1971, doi: 10.1016/S0010-2180(71)80012-3.

[23] T. Saito and I.I. Glass, "Temperature measurements at an implosion focus," *Proceedings of the Royal Society of London. A. Mathematical and Physical Sciences*, vol. 384, no. 1786, pp. 217–231, Nov. 1982, doi: 10.1098/RSPA.1982.0156.

[24] H. Matsuo, K. Ebihara, Y. Ohya, and H. Sanematsu, "Spectroscopic study of cylindrically converging shock waves," *J Appl Phys*, vol. 58, no. 7, pp. 2487–2491, 1985, doi: 10.1063/1.335925.

[25] M. Kjellander, N. Tillmark, and N. Apazidis, "Thermal radiation from a converging shock implosion," *Physics of Fluids*, vol. 22, no. 4, p. 046102, Apr. 2010, doi: 10.1063/1.3392769.

[26] C. O. Laux, M. Winter, J. Merrifield, A. Smith, and P. Tran, "Influence of ablation products on the radiation at the surface of a blunt hypersonic vehicle at 10 km/s," *41st AIAA Thermophysics Conference*, 2009, doi: 10.2514/6.2009-3925.

[27] V.S. Saranyamol, N. Soumya Ranjan, and S. Mohammed Ibrahim, "On spherical shock wave focusing in air — A computational study," *European Journal of Mechanics - B/Fluids*, vol. 91, pp. 27–37, Jan. 2022, doi: 10.1016/J.EUROMECHFLU.2021.09.009.

[28] A. M. Brandis and B. A. Cruden, "Shock tube radiation measurements in nitrogen," *2018 Joint Thermophysics and Heat Transfer Conference*, 2018, doi: 10.2514/6.2018-3437.

[29] A. B. Meinel, "The spectrum of the airglow and the aurora," *Reports on Progress in Physics*.

[30] M. R. Gochitashvili, R. Y. Kezerashvili, and R. A. Lomsadze, "Excitation of Meinel and the first negative band system at the collision of electrons and protons with the nitrogen molecule," *Phys Rev A*, vol. 82, no. 2, p. 022702, Aug. 2010, doi: 10.1103/PHYSREVA.82.022702/FIGURES/6/MEDIUM.